\documentclass{egpubl}
\usepackage{eg2023}
\usepackage{textcomp}
\usepackage{booktabs}

\ShortPresentation      
\usepackage[T1]{fontenc}
\usepackage{dfadobe}
\usepackage[caption=false]{subfig}
\usepackage{tabulary}

\usepackage{cite}  
\BibtexOrBiblatex

\electronicVersion
\PrintedOrElectronic
\ifpdf \usepackage[pdftex]{graphicx} \pdfcompresslevel=9
\else \usepackage[dvips]{graphicx} \fi
\usepackage{xcolor}
\usepackage{tikz}
\usetikzlibrary{arrows.meta}
\tikzset{%
  >={Latex[width=5mm,length=5mm]},
            base/.style = {rectangle, rounded corners, draw=black,
                           minimum width=6cm, minimum height=1.5cm,
                           text centered},
  activityStarts/.style = {base, fill=blue!20},
       startstop/.style = {base, fill=red!20},
    activityRuns/.style = {base, fill=green!19},
         process/.style = {base, minimum width=2.5cm, fill=orange!15
                           },
}

\usepackage{egweblnk} 

\title {Velocity-Based LOD Reduction in Virtual Reality: A Psychometric Approach}

\author[D. Fellner \& S. Behnke]
{\parbox{\textwidth}{\centering David Petrescu\orcid{0000-0002-8949-7265}, Paul A. Warren\orcid{0000-0002-4071-7650},
        Zahra Montazeri\orcid{0000-0002-2091-8535} and
        Stephen Pettifer\orcid{0000-0002-1809-5621}
        }
        \\
{\parbox{\textwidth}{\centering University of Manchester\\
       }
}
}

\begin{document}

 \teaser{

}
\maketitle

\begin{abstract}
  
 

Virtual Reality headsets enable users to explore the environment by performing self-induced movements. The retinal velocity produced by such motion reduces the visual system's ability to resolve fine detail. We measured the impact of self-induced head rotations on the ability to detect quality changes of a realistic 3D model in an immersive virtual reality environment. We varied the Level-of-Detail (LOD) as a function of rotational head velocity with different degrees of severity. Using a psychophysical method, we asked 17 participants to identify which of the two presented intervals contained the higher quality model under two different maximum velocity conditions. After fitting psychometric functions to data relating the percentage of correct responses to the aggressiveness of LOD manipulations, we identified the threshold severity for which participants could reliably (75\%) detect the lower LOD model. Participants accepted an approximately four-fold LOD reduction even in the low maximum velocity condition without a significant impact on perceived quality, which suggests that there is considerable potential for optimisation when users are moving (increased range of perceptual uncertainty). Moreover, LOD could be degraded significantly more in the maximum head velocity condition, suggesting these effects are indeed speed dependent.

\ccsdesc[500]{Computing methodologies~Perception}
\ccsdesc[300]{Computing methodologies~Virtual reality}

\printccsdesc   
\end{abstract}  

\section{Introduction}
Recent growth in the availability of high-performance graphics technology and consumer-grade Head Mounted Displays (HMDs) has led to an explosion in the use of Virtual Reality (VR), which is now commonplace in research, leisure and commercial contexts. Generating compelling, fully-immersive VR experiences, however, is notoriously computationally expensive, with expectations typically outstripping capability. In spite of advances in CPU and GPU power, there is a clear need for the development of techniques and algorithms that can provide the best user-experience for a given level of raw hardware performance. One approach to doing this focuses on exploiting characteristics and limitations of human perceptual systems. Here we focus on changes in human sensitivity that relate to  movement of the observer, and therefore lead to global patterns of motion on the retina. 

When we rotate our heads (or eyes) in order to track a moving object or to attend to a salient stimulus, our visual sensitivity to background elements is thought to be decreased \cite{murphy_pattern_1978}. For example, VR users often produce fast rotational head or eye movements when tracking a cross-hair or an object, which creates a pattern of motion across the whole retina. Building upon this phenomenon and previous work that considers velocity-based Level-of-Detail (LOD) reduction methods, we introduce a psychometric approach to determine how observer movement influences the sensitivity to a drop in the quality of a background stimulus.


Most HMDs allow tracking of user movement with 6 Degrees of Freedom (6DOF) and visual representations of stimuli that accompany such movement. This enables users to move freely in the virtual environment and provides access to proprioceptive and vestibular information from beyond the retina about such movements, information which is thought to be important for the correct interpretation of a stable scene during self movement \cite{warren_investigating_2022}. However, there is also evidence that the Human Visual System's (HVS) sensitivity is altered when there is global motion on the retina resulting from observer eye/head movements \cite{braun_visual_2017}. The relationship between movement and sensitivity is complex, depending on visual features in the display and the types of movement undertaken. To the best of our knowledge, such results are not well-integrated in current approaches to optimising VR environment rendering. In light of this we will focus on two important questions:
\begin{itemize}
    \item Might the reduction in visual sensitivity during observer movement allow for a speed dependent reduction in the complexity of a rendered virtual model without it being perceived by the user?
    \item To what extent can the complexity be reduced before being perceived?
    
\end{itemize}


In this paper we address these questions by exploring the feasibility of a velocity-dependent scene-complexity reduction algorithm in a commercial VR HMD. The geometrical LOD represents the level of complexity of a virtual model. Generally, models have multiple LODs at different quality levels. The levels are generated by simplifying the mesh complexity of the initial model. In most commercial game engines, LOD changes occur when the distance of the object from the view plane reaches a certain threshold as the spatial acuity of the eye is not sensitive enough to notice differences, but they do not account for other perceptual factors. In recent years, perceptually-guided LOD systems have been studied relatively sparsely in contrast with other approaches to the problem that focus on on real-time physical based shading and the need for perceptual metrics for sampling optimisation, but still represent a fruitful area to explore. 

Therefore, the contributions of this paper are: 
\begin{itemize}
     \item We test the feasibility of a lightweight method that only uses velocity as a metric for guiding LOD degradation (i.e. agnostic to other properties of the HVS). 
    \item Using psychophysical methods, we clearly demonstrate that self-induced horizontal head rotations allow a significant (approximately four-fold) degradation in LOD for a well-lit model.
    \item We provide evidence that the level of degradation tolerated by users increases for faster movements, as predicted.
    
\end{itemize}




\section{Related Work}\label{section:related}
\subsection{Spatio-temporal considerations and perceptual rendering}
Considerable work has been done in investigating how the limitations of the HVS can be used to guide algorithms by detecting potential rendering artifacts. Many of these vision models are based on early psychophysical research that investigated the Contrast Sensitivity Function (CSF) under various spatial and temporal conditions \cite{robson_spatial_1966}. The CSF is widely used in various forms to speed-up the rendering of images and animations by creating metrics that predict what the HVS will perceive. For example, Daly's Visible Difference Predictor (VDP) \cite{daly_visible_1992} uses the frequency domain of two images to predict where changes in contrast become unnoticeable. This has been used in many applications in which a reference frame is available to stop unnecessary rendering that would not induce a meaningful perceptual change. Kelly explored a spatio-velocity model of the CSF by analysing sinusoidal gratings moving at various velocities in order to determine our sensitivity thresholds \cite{kelly_motion_1979}. Daly later improved on the model by also accounting for eye movements \cite{daly_engineering_1998}. Myzskowski et al. used a spatio-velocity approach to develop the Animation Quality Metric (AQM) \cite{myszkowski_perceptually-informed_1999} which improved the rendering quality of perceptually important frames (termed "keyframes") which could affect the overall quality percept of dynamic scenes because of certain properties easily noticeable by the HVS. Yee et al. also created a comprehensive model which contained a spatio-temporal error map by using motion estimation and attention models \cite{yee_spatiotemporal_2001}. Recently, Mantiuk et al. used data from multiple papers in order to model the CSF over 5 dimensions (Spatial, Temporal, Eccentricity, Luminance, Area) and named it stelaCSF \cite{mantiuk_stelacsf_2022}. Models that predict the quality of motion in a velocity-dependent way account for movement artifacts \cite{denes_perceptual_2020}, and others that use the Variable Rate Shading \cite{nvidia_vrworks_2018} to model masking of degradation artifacts due to motion \cite{yang_visually_2019} and to model local shading and refresh rate\cite{jindal_perceptual_2021}. During horizontal head rotations such as those presented in our paper, all the dimensions present in the stelaCSF are relevant, but they do not account for the type of user movement. Lavou\'{e} created a metric for generating distortion maps for detecting differences between geometrically distorted and reference meshes that can be used to guide optimisation and evaluation. This metric is highly correlated with the mechanism used by the HVS when perceiving such changes \cite{lavoue_multiscale_2011}. Applications that consider vestibular factors are less common. Ellis et al. have investigated the cross-modal interactions between the vestibular response and the visual system in both translational and rotational movements
\cite{ellis_effect_2006,ellis_effect_2006-1}. They used a `pod'-like 6DOF Motion Simulator, with participants sat surrounded by a display. A central area on the screen was rendered and sampled at high-quality (HQ). When the pod was performing translation movements, essentially moving towards the HQ region, the area would decrease in size. Using biological models of the vestibular system, the area HQ region was decreased as a function of the acceleration of the pod. During rotation, they found some scope for improvement, but more fine-tuning was required for a significant effect. Our paper is related to this work in identifying such parameters during self-imposed movements.

\subsection{Adaptive Level-of-Detail}
Adaptive LOD methods have been used extensively in the past. Generating LODs causes geometrical changes and irregularities. Algorithms were therefore refined to accommodate these factors. Solutions vary from accounting for collapsing edges or pairs of vertices using quadric matrices to preserve surface error \cite{garland_surface_1997}, to using dynamic algorithms such as those proposed by Hoppe \cite{hoppe_progressive_1996}. He introduced progressive meshes, which allow for smoother and reversible transitions at no extra memory cost, using only the highest quality mesh available for computing lower quality transformations. View-dependent strategies for efficient LOD selection have been used before; Hoppe combined his progressive LOD generation with view-dependent models to improve on his progressive mesh generation \cite{hoppe_view-dependent_1997}. Luebke and Erikson lowered LODs by generating a hierarchy of vertex clusters potentially visible from the current view point \cite{luebke_view-dependent_1997}. Watson et al. investigated how performance in a search task was affected by degrading the LOD in the peripheral region of the gaze \cite{watson_managing_1997}. They found that participants tolerated an almost two-fold LOD degradation before performance was affected. Similarly, Murphy and Duchowski used a gaze-contingent method in a multi-resolution mesh by degrading parts of the model \cite{murphy_perceptual_2002}. Howlett et al. used gaze-contingent methods to identify salient features of objects that are likely to be attended to by viewers \cite{howlett_experimental_2004}. This meant that higher resolution meshes were only used in areas that were perceptually important. Other methods looked at LOD reduction in various human animations or collision events and used psychometric testing for data collection \cite{osullivan_collisions_2001}. Velocity-based LOD reduction has not traditionally receive as much attention. Funkhouser created a heuristic-based quality reduction model; the apparent speed of the object was used as a heuristic because of the inability to discriminate detail when motion-blur at high speeds was present \cite{funkhouser_adaptive_1993}. Reddy created a comprehensive model for velocity-based LOD selection taking into account eccentricity and velocity for moving objects across the visual field \cite{reddy_specification_1998,reddy_perceptually_2001}. Parkhurst and Niebur used a search task to test the feasibility of a velocity-based LOD reduction system that considered a progressive LOD reduction proportional to the speed of the camera \cite{parkhurst_feasibility_2004}. The task-object decreased and increased in quality based on the speed of the viewport. They found that the computational savings generated by such methods often increase the stability of the environment and behavioural consequences, albeit while impeding task performance. Building on these methods, we are interested in studying how self-generated motion in a modern VR HMD affects the percept of quality during LOD changes. Our paper introduces a psychometric approach to  measure the perceptual sensitivity to LOD changes during ego-motion. Thus, we propose a rigorous framework for measurement of thresholds as participants are making direct judgements about quality changes.

\section{Methods}
In the present experiment, over a series of trials, participants were asked to track a fixation point by performing yaw rotations of the head while judging the quality of two models presented in different intervals. One of the intervals always contained the reference model for the duration of the interval and the other contained the same model, but with LOD degradation (varying in severity from trial to trial) applied whenever rotational head velocity exceeded a given threshold. The participant's task was to identify the interval containing the degraded model. In perceptual psychophysics this is referred to as a two-interval forced choice (2-IFC) procedure and such methods are used extensively to recover estimates of perceptual thresholds \cite{warren_investigating_2022}. 

\subsection{Participants}
Seventeen participants (all students or staff at the University of Manchester, fifteen were naive to the purpose of the study whereas 2 had expertise in using VR) were recruited. The study was approved by the Ethics Committee of The University of Manchester, Division of Neuroscience and Experimental Psychology. Informed consent was given by all participants. Two participants were excluded from the analysis since their responses were particularly noisy and it was not possible to fit summary parameters to their data (see below - Section \ref{section:PF} for information on the the fitting procedure).

\subsection{Apparatus}
The data for this experiment were collected in the Virtual Reality Research (VR2) Facility at the University of Manchester. The HTC Vive HMD was used with a wireless receiver to display tethered content from the PC (Intel i7-7700K CPU and NVIDIA RTX 3080 Ti GPU). The HTC Vive has a refresh rate of 90 Hz and a resolution of 1080 x 1200 pixels for each eye. Using our PC setup we kept the frame rate at approximately 90 FPS which matches the refresh rate of the HMD and thus minimised frame drops that could cause nausea or any other discomfort.

The experiment was coded using the \textit{Unity} game engine and Unity Experiment Framework (UXF), which was designed for the development and control of psychophysics studies \cite{brookes_studying_2020}. Participants used the trigger on either the left or right Vive controller to start the experimental trials and make responses. 

\subsection{Design}
In order to measure the impact of head speed on the participants' ability to perceive LOD changes, we manipulated maximum user head \textit{speed} as an independent variable with two levels (\textit{Fast} and \textit{Slow}). We used a fully within-participants design, meaning all participants took part in both conditions. For each of the two speed conditions, participants undertook 280 trials for which the severity of the LOD degradation applied (here termed the `aggressiveness') was varied over a range of seven pre-determined values (see \textit{Stimulus} section for more details). For each aggressiveness level participants repeated the same number of trials. This approach is commonly referred to as the Method of Constant Stimuli (MCS) in the perceptual psychophysics literature \cite{simpson_method_1988}.


\subsection{Stimuli}
\begin{figure*}[!hbt]
\centering
\hspace{-2cm}
\includegraphics[width=1.08\linewidth]{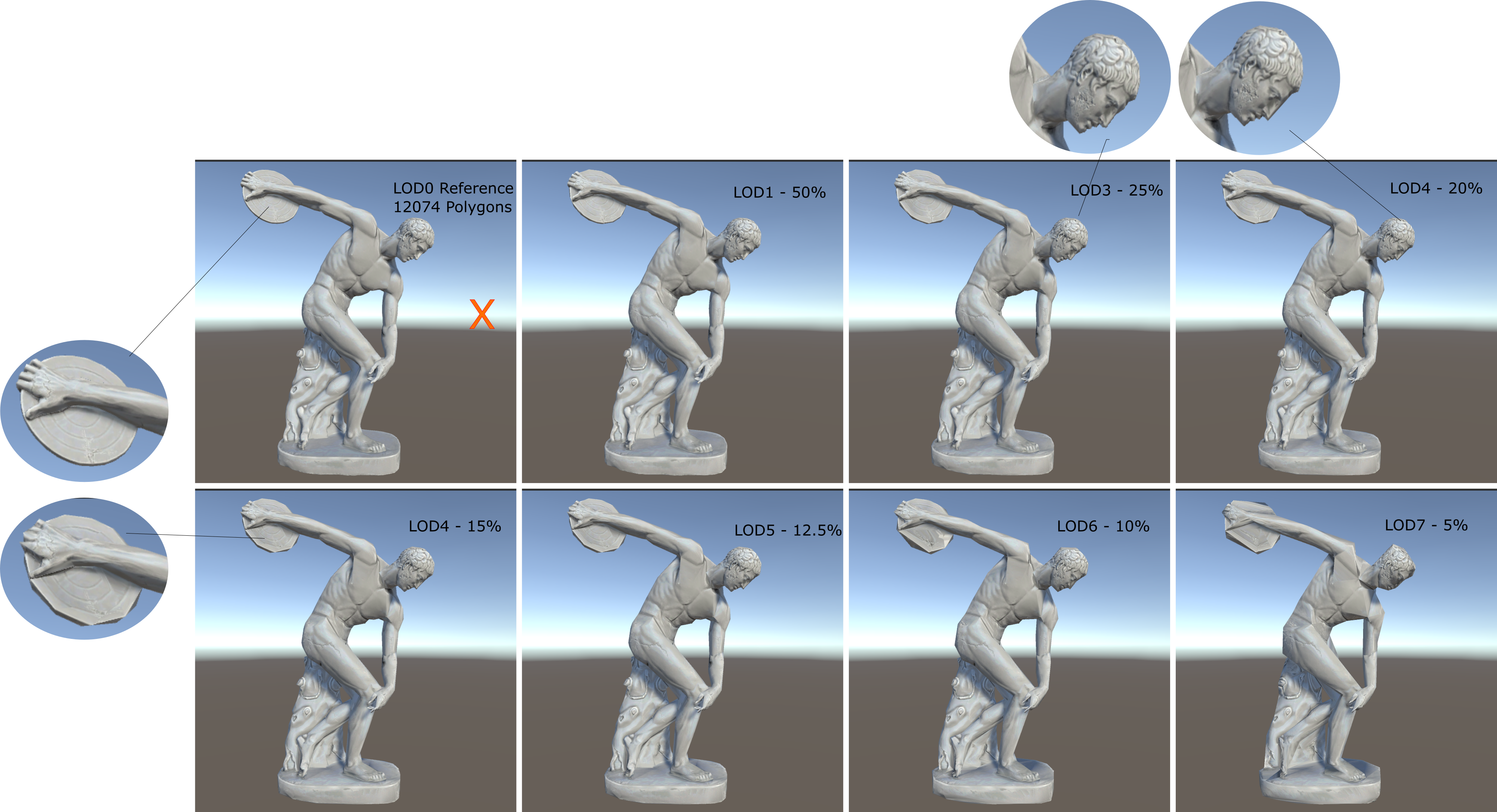}
\caption{LOD levels generated using the algorithm in \cite{edlund_unitymeshsimplifier_2022}}
\label{fig:LODs}
\end{figure*}

\begin{figure}[tbp]

  \centering

  \includegraphics[width=1\linewidth, height=0.5\linewidth]{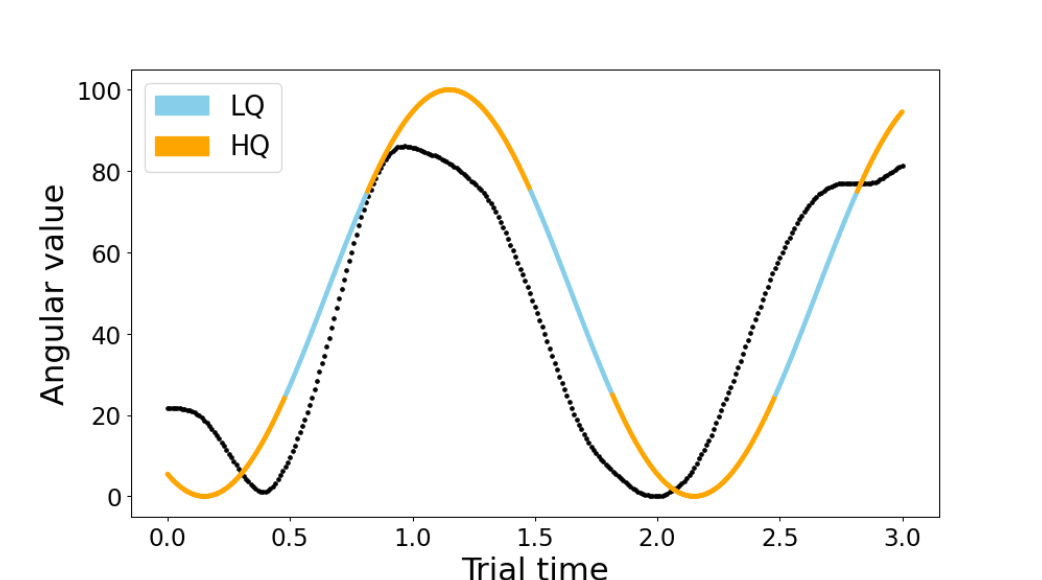}
  \caption{\label{fig:motion}%
           The angular position of a participant's head sampled during the trial (black) mapped against the angular position of the fixation target. Points at which the stimulus changed LOD are highlighted in orange and blue.}
\end{figure}

A moving fixation cross (which participants were asked to track) was presented on each trial (present in the top left portion of Figure \ref{fig:LODs}).The cross occupied around 2x2 degrees of the visual field. The cross was color-coded to differentiate between the two different presentation intervals using colours discriminable by colour blind participants. The fixation point moved in the virtual environment on a fixed radius circular trajectory centred on the participant's head and with a range of 100\textdegree (i.e. 50\textdegree to the left and right of straight ahead). The radius was given by the distance between the participant and the statue, which was roughly 5 meters in virtual space. The angular velocity of the fixation cross (relative to the head) was sinusoidal - Figure (\ref{fig:motion}). In the \textit{Fast} condition, the velocity peaked at approximately 157\textdegree/sec and in the slow condition at around 52\textdegree/sec. During head movements, the vestibulo-ocular reflex (VOR) stabilizes the eye, so the retinal image of the object is more accurate \cite{pulaski_behavior_1981}. Although some authors assume peak velocities of up to 900\textdegree/sec \cite{kothari_gaze--wild_2020}, there is evidence that the VOR is stable around 170\textdegree/sec \cite{grossman_frequency_1988} and that in visual search tasks the VOR reflex is reliable until 350\textdegree/sec \cite{pulaski_behavior_1981}. 

The primary stimulus was an environment-stationary model of the ancient Greek statue \textit{Discobolus} by Myron, which is a free asset from \textit{ChamferBox Studio} on the Unity Store \cite{chamferbox_discobolus_2018}. The model was presented in the default Unity scene in a well-lit environment (see Figure \ref{fig:LODs}).

\begin{figure}
    \centering
    \scalebox{0.6}{
    \begin{tikzpicture}[node distance=1.5cm,
    every node/.style={fill=white, font=\Large}, align=left]
  \node (start)             [activityStarts]              {Run program};
  \node (onCreateBlock)     [process, below of=start,yshift = - 1cm]          {Generate LODs};
  \node (onStartBlock)      [process, below of=onCreateBlock,yshift = - 1cm]   {Pick LOD};

  \node (activityRuns)      [activityRuns, below of=onStartBlock,yshift = - 1cm]
                                                      {Render LOD};
  \node (CalcVelBlock) [startstop, below of=activityRuns, yshift = - 2cm]{Record Velocity};

  \node (onRestartBlock)    [process, right of=onStartBlock, xshift=4cm]
                                                              {Map Vel. Values};
  
  \draw[->]             (start) -- (onCreateBlock);
  \draw[->]     (onCreateBlock) -- (onStartBlock);
  \draw[->]     (onStartBlock) -- (activityRuns);
  \draw[->]      (activityRuns) -- node[text width=5.5cm]
                                   {Record and calculate velocity} (CalcVelBlock);
 
  \draw[->]      (CalcVelBlock) -| node[yshift=1.25cm, text width=4cm]
                                   {Check if velocity threshold was reached}
                                   (onRestartBlock);
  \draw[->]     (onRestartBlock) -- (onStartBlock);

  \end{tikzpicture}
}
    \caption{Workflow of our method. After the activity starts, our lightweight method generates LODs based on the user input. The base LOD is picked before any velocity is recorded. After the rendering process, velocity is recorded and mapped to an LOD map. This is done via normalising the angular velocity recorded to an interval of speeds where 0 is stationary and the max value is the expected maximum velocity (in this case given by the fixation point).}
    \label{fig:workflow}
\end{figure}
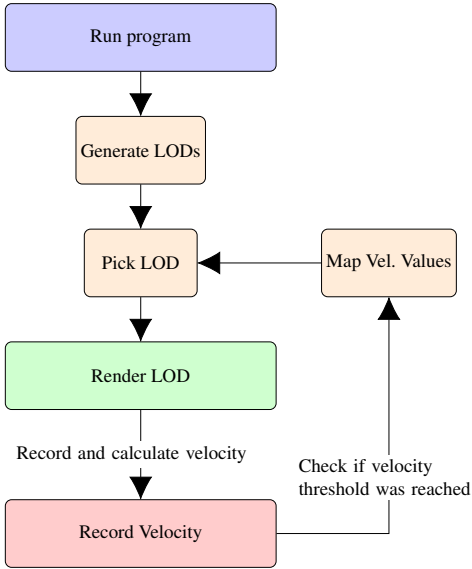

The LOD degradation process applied to the statue model was an adjusted version of the quadric metric mesh simplification algorithm introduced by Garland and Heckbert \cite{garland_surface_1997}. This algorithm is a fast quadric edge-collapse written in C\# that can be used as a Unity asset \cite{edlund_unitymeshsimplifier_2022} and was adapted from the original version written in C++ \cite{spacerat_fast-quadric-mesh-simplification_2022}. The workflow we used for our method can be seen in Figure \ref{fig:workflow}. The algorithm acts to reduce the number of polygons of the mesh and was adapted to enable speed-dependent application such that the LOD degradation is implemented once a minimum threshold user head speed was crossed. Degradation was applied at one of seven levels of aggressiveness detailed in Figure \ref{fig:LODs}. The reference model was composed of 12074 polygons and in the most aggressive LOD degradation condition this was reduced by a factor of 20 to around 600 polygons. When the rotational velocity of the head reached 50\% of the peak velocity of the fixation point, the quality of the statue was degraded. This meant that quality of the statue was reduced for around half of the number of frames for which the participant moved their head. As seen in Figure \ref{fig:motion}, by plotting the horizontal angular displacement of a participant during a trial, the sinusoidal velocity trend of the fixation point is matched, which gave us confidence that the intended head rotational velocity was achieved in the trials.

\subsection{Procedure}
Participants were instructed to put on the HMD and helped to adjust the tightness and lens distance for comfort and to provide a crisp image. Participants who were unfamiliar with VR first spent several minutes in the HTC Vive home environment and were monitored for dizziness or nausea. After the task was explained to them, participants were given the opportunity to undertake several practice trials to ensure they understood the task and were familiar with the response buttons. The participant's position was then aligned with the mid-sagittal plane of the statue. 

Each trial began with a participant fixating on the stationary fixation cross. The participant initiated movement at their own pace by pressing one of the buttons on the hand-held controller. The initial position of the cross was randomised in the 100 degree range across trials but the participant was cued (by displacing the fixation point to the either side of statue) to the direction in which the cross would begin to move. The fixation cross then moved for 2.5 seconds per interval. The participant was then cued to the start of the second interval, with another stationary fixation cross before it moved again for another 2.5 seconds. The reference model was presented on every trial in either the first or second interval at random and one of the other LOD models was presented in the other interval. The participant was instructed to indicate which interval (first or second) contained the higher-quality model. Each trial lasted around 5 seconds plus the time the participant took to start the intervals and record the answer. 

Over the course of the experiment each participant provided 20 responses for each of the 7 possible pairs of reference and LOD models. In addition we interleaved trials for the fast and slow maximum head velocity conditions. Consequently, in total participants provided 280 responses. In order to reduce participant fatigue, we divided the experiment into 2 sessions, each lasting around 40 minutes.



\subsection{Psychometric Function Fitting}\label{section:PF}

\begin{figure}[tbp]
    \centering
    \includegraphics[width=1\linewidth]{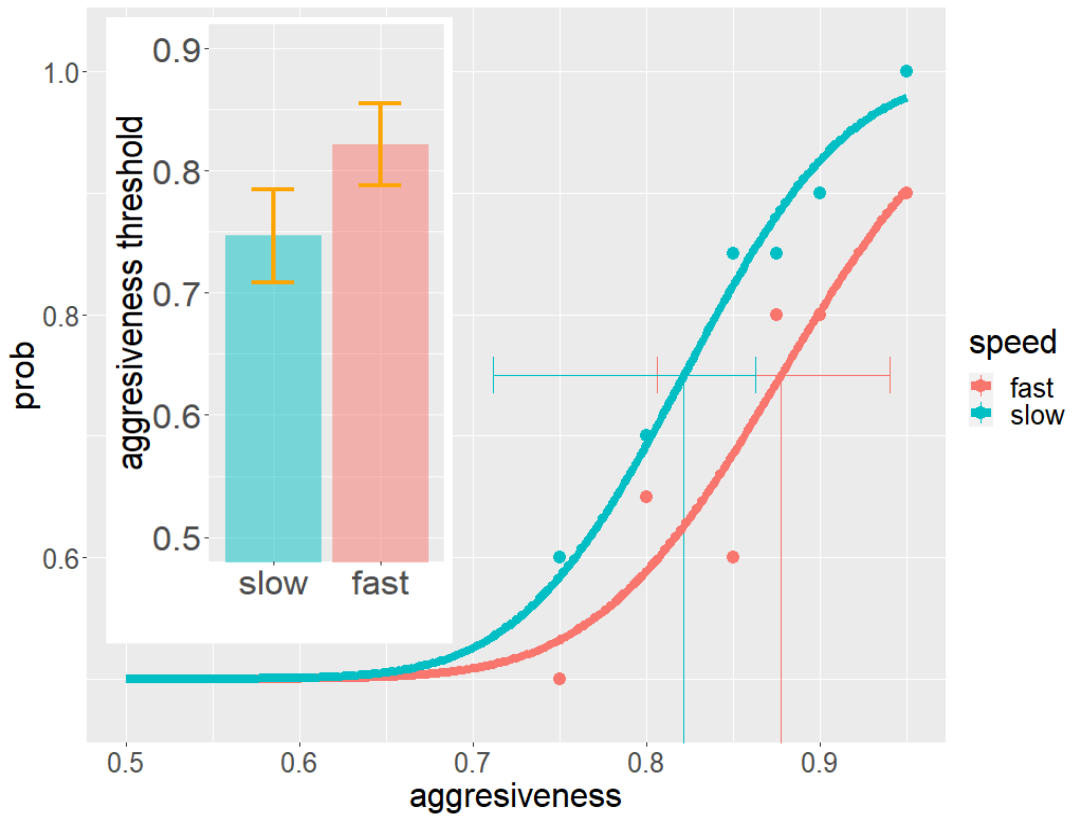}
    \setlength{\belowcaptionskip}{-18pt}
    \caption{Bar plot showing the mean threshold values at the 75\% point (Left - Embedded). Example psychometric functions recovered from one Participant. \textit{Fast} - in red; \textit{Slow} - in blue}
    \label{fig:PF}
\end{figure}

To analyse the data for each participant and for each of the two maximum head velocity conditions, we first plotted the LOD values against the proportion of times (out of 20 repeated trials) that the participant correctly identified the interval containing the degraded statue model. The proportions obtained should vary between 50\% (chance performance) for a small LOD degradation, up to 100\% (perfect performance) for very large levels of degradation, and follow an ogive shape reflecting the accrual of more evidence as LOD degradation severity increases. We modelled this data as a psychometric function (PF). A psychometric function represents a model of how observer perception (here the percept of which model has higher quality) changes as a function of some physical parameter (here the LOD manipulation). We used a cumulative Gaussian with two free parameters (mean and standard deviation) as the psychometric function model. Fits to the data using this model were obtained using the \textit{quickpsy} package \cite{linares_quickpsy_2016} in R, which is designed to undertake maximum likelihood psychometric function fitting to data of this form. Example data and fits for two observers are illustrated in Figure \ref{fig:PF}. From these fits we then recovered the threshold LOD value which resulted in 75\% performance (vertical lines in Figure \ref{fig:PF}). This threshold parameter reflects the LOD value for which participants can reliably (75\% of the time) distinguish the reference from the degraded stimulus. Poorer ability to reliably distinguish the reference from test interval will be reflected in higher threshold values. Our hypotheses would therefore be confirmed if the thresholds observed are high and increase as a function of maximum head velocity.

\section{Results}
Based on the function fitting above, we recovered statistics for each participant in each of the conditions. The embedded graph in Figure \ref{fig:PF} illustrates the mean threshold over the 15 participants in this experiment for the \textit{Slow} and \textit{Fast} maximum head velocity conditions. Note that even in the \textit{Slow} condition the LOD degradation tolerated by participants before they can reliably identify the interval containing the degraded model is around 75\%. This corresponds to an 4-fold reduction in the number of polygons. Note also that this threshold appears to be higher for the high maximum head velocity condition. The threshold in the \textit{Fast} condition was around 82\%. We ran a 1-tailed paired t-test on these data and found that the threshold obtained in the \textit{Fast} condition was significantly higher than in the the \textit{Slow} condition (t = 2.71, \textit{p} = 0.008), consistent with our hypothesis that the effects are speed dependent. 


Taken together these data are consistent with our hypotheses in suggesting that:
\begin{itemize}
    \item Humans are markedly insensitive to large degradation in geometrical detail (i.e. polygon count in our experiments) during head movements
    \item They become more insensitive as speed of head movement increases
\end{itemize}

These findings suggest that there is potential for a simple optimisation approach if quality of rendering is made dependent on head (and/or eye) movement speed.

\begin{table}[!hbt]
\centering
\begin{tabular}{ccccc}
\hline 
N=15           & \textit{slow}    & \textit{fast}    & \textit{t} & p              \\ \hline 
\textbf{LOD\%} & 74.6 (s.d. 14.8) & 82.2 (s.d. 13.1) & 2.71       & \textless 0.01 \\ \hline
\end{tabular}
\caption{Summary statistics (mean and s.d.) for two conditions and outcome of paired t-test}
\label{tabel:resultsTable}
\end{table}

\section{Discussion}

Our method provides a different approach to switching between LOD levels that is independent of the distance between the camera and the objects in the scene (the standard approach in most commercial game engines). We have shown that even with very aggressive quality degradation participants cannot reliably perceive these changes.

Nonetheless, there is clearly scope to achieve even greater degradation in LOD without a perceptual consequence if we consider additional higher level cognitive processes (e.g. attention, decision making, etc.). Combining our LOD degradation method with gaze-contingent methods such as those presented by Watson and Duchovski \cite{watson_managing_1997} could lead to an aggregated result similar to that achieved by Reddy \cite{reddy_specification_1998}.  These results are consistent with the findings of Parkhurst and Niebur \cite{parkhurst_feasibility_2004}, in which a model was degraded by a factor of more than 3 without an impact on task performance in a search task. This suggests that perceptual LOD degradation is still a valuable asset to explore and that this is the case even when participants are actively looking for a degradation.

\textbf{Other types of user movement.} Note that although we used head movement to generate the retinal motion, we suggest that these results are likely to hold irrespective of the cause of the movement input. More specifically we would expect the same effects to be observed if the participant had been instructed to follow the target by making smooth pursuit eye movements (SPEMs). SPEMs allow the HVS to stabilize the tracking of a moving target in order to hold the target in the most sensitive foveal region of the retina. Alternatively we also suggest that we would have obtained similar results if participants had been asked to fixate on a stationary cross and the statue had moved in a manner consistent with the motion arising from the head movement we examined in our experiment. While we have not tested these possibilities it seems likely that the key reason for the insensitivity to the degradation is the motion itself rather than the cause of that motion, something Murphy pointed out in his seminal work \cite{murphy_pattern_1978}. In particular with respect to SPEMs, Meyer \cite{meyer_upper_1985}, suggests that these are reliable up to approximately 100\textdegree/sec at which point tracking performance starts to decline and other mechanisms (e.g. simultaneous eye and head movements) are required to aid with tracking. Other work points out that SPEM does cause modulations in the visual acuity of the user under variations of background spatial frequency or luminance changes \cite{schroder_replicability_2021,braun_visual_2017}. In our case, the fixation point moved at 57\textdegree/sec for the \textit{Slow} condition and 157\textdegree/sec respectively for the \textit{Fast} condition so it is likely that the tracking involved both head and to a lesser-extent eye movements and the VOR system which acts to stabilise targets on the retina \cite{pulaski_behavior_1981}. This work speaks to the need for more in-depth analysis of different causes of observer movement and, potentially, across a wider range of velocities. Metrics recovered from such analyses could be useful for inclusion in vision models such as those presented in \cite{mantiuk_stelacsf_2022-1,lavoue_multiscale_2011,yee_spatiotemporal_2001}.

\textbf{Limitations and Future Work.} Our method has several limitations. First, because we used a discrete LOD change method; occasionally this would introduce detectable flicker in the degraded model. Of course detection of such flicker could give participants a cue as to which interval contained the degraded model and, if it were eliminated, the LOD degradation tolerated by users could have been even more severe. Flicker might be removed in future studies by using smoothing techniques or progressive meshes, such as the ones introduced by Hoppe \cite{hoppe_progressive_1996}. 

A second issue is that we did not track the eyes and so could not be sure that participants were following instructions perfectly. Nonetheless, the head movement observed was generally close to the instructed movement (see Figure \ref{fig:motion}). 

Third, we examined only one type of observer movement. Clearly we might find subtly different effects for e.g. different directions of movement, different combinations of eye, head and model movement or different speeds of movement. To test all these was not feasible in our study but, clearly, could be the basis of future work.  

Fourth, there may be less benefit with using this method for people who are more experienced with VR. This may be due to the so-called `Expertise Effect' \cite{brajnik_expertise_2011}. This states that experienced users tend to detect "problems" with certain systems more reliably than inexperienced ones. This was also true for our 2 non-naive participants, as the threshold LOD was lower than that for the average.

In addition to the issues raised above, future work should consider other quality aspects of models and how they are influenced by movement. For example, one avenue for potential study would focus on how perception of complex (and computationally demanding) material properties are altered by retinal motion (either due to movement of the user or the material itself. Mao et al. investigated the role of motion in the correct perception of material properties \cite{mao_effect_2019}. In doing so, they added different degrees of motion blur to a set of stimuli with various material properties and found that the perception of certain attributes are significantly affected by motion. We intend to extend our framework to study the impact of self-induced motion on material properties. 

\section{Conclusion}

In this paper we re-examine the idea of velocity-based rendering using a psychophysical approach to measure precisely the extent to which users can reliably detect degradation of a model in the scene in the presence of retinal motion. Our method differs from previous approaches because: i) model degradation rests upon information about the self-induced velocity of movement (and ignores distance in the scene or information about object velocities in the scene); ii) it examines sensitivity to quality degradation when users are directly instructed to detect such changes; iii) it uses well-established psychophysical methods to provide precise estimates of the parameters in question.  

 Our data suggest that even in slow head rotations a degradation of up to 75\% relative to the original model is accepted by participants, even when they are asked to pay attention to it directly. We also confirmed that higher rotational velocities produce a significantly larger effect. Given that information about the position and movement of a tracked HMD is easily accessible in most game engines this direct approach to LOD reduction could be useful in future.

\bibliographystyle{eg-alpha-doi} 
\bibliography{DavidPhD}  

\end{document}